# Multibranched parametric resonance and swallowtail catastrophe in electromechanical oscillators with nonlinear friction

P. Y. Chan[1], L. Huang[1], X. Dong[1], M. I. Dykman[4] and H. B. Chan[1,2,3,*]

[1]Department of Physics, the Hong Kong University of Science and Technology, Clear Water Bay, Kowloon, Hong Kong, China.

[2] William Mong Institute of Nano Science and Technology, the Hong Kong University of Science and Technology, Clear Water Bay, Kowloon, Hong Kong, China.

[3] IAS Center for Quantum Matter, the Hong Kong University of Science and Technology, Clear Water Bay, Kowloon, Hong Kong, China.

[4] Department of Physics and Astronomy, Michigan State University, East Lansing, Michigan 48824, USA.

*Author to whom any correspondence should be addressed.

E-mail: hochan@ust.hk

Keywords: parametric oscillators, nanomechanical resonators, nonlinear friction, swallowtail catastrophe

## Abstract

Parametric resonance underpins the operation of a wide range of physical systems, from nanomechanical resonators to quantum-information systems and Ising machines. As an archetypal class of driven-dissipative systems, parametric oscillators are generally expected to exhibit a single pair of stable period-two states with opposite phases. This bistable behavior enables both the simulation of spin Hamiltonians and the preparation of superconducting cat states. Whether multiple pairs of such states can coexist in a single oscillator, however, remains an open question. Here, we show experimentally and theoretically that conventional controlled nonlinear friction can induce the coexistence of two distinct pairs of period-two states in a micromechanical oscillator. The friction is implemented via a canonical approach, utilizing a drive-induced resonant coupling that transfers two vibrational quanta from the oscillatory mode to a faster decaying mode. We demonstrate that the onset of multistability is governed by a swallowtail catastrophe and quantitatively map the associated bifurcation structure. Our results broaden the understanding of parametric resonance and establish micro- and nano-mechanical oscillators as a versatile platform for studying catastrophe theory and multistable nonequilibrium dynamics.

## 1. Introduction

Nonlinear dissipation is characterized by an energy decay rate that depends on the oscillation amplitude [1,2]. It has long served as an essential mechanism for stabilizing limit-cycle oscillations, most notably within the Van der Pol framework, starting with radio-frequency generators [3]. Recently, there has been a resurgence of interests in nonlinear dissipation. Conventional nonlinear dissipation arises from nonlinear coupling to a broadband bosonic bath [4]. In nano-mechanical systems, nonlinear friction typically comes from nonlinear phonon-phonon interactions [5,6] or anchor losses [7]. The decay process involves the removal of two vibrational quanta from the oscillatory mode to the thermal reservoir. It can give rise to amplitude-dependent linewidths and decay rates [8-20]. Applications of nonlinear friction are expanded by the possibility to control it using resonant driving that nonlinearly couples the mode of interest to a mode with a higher decay rate [15,20].

Nonlinear friction plays a particularly important role in parametric oscillators. Such oscillators have been attracting much interest recently in various contexts [21]. In particular, coupled parametric oscillators show promise in creating Ising machines [22-27] and in modeling asymmetric Ising networks [28]. In the field of quantum information, parametric oscillators with nonlinear dissipation offer a major platform for implementing cat qubit [29-34]. Under the conventional model of nonlinear friction, the energy decay rate of an oscillator increases monotonically with oscillation amplitude, which provides an important mechanism of suppressing the unbounded increase in the amplitude of parametric oscillators [35,36].

Given this broad interest, it is important to fully characterize the dynamics of parametric oscillators with controlled nonlinear friction. However, so far, neither the number of period-two states that can coexist nor the range of parameter values supporting the coexisting states have been established. Since the modes of both superconducting microwave resonators and nano- and micro-mechanical resonators are strongly underdamped, parametric resonance emerges where the parametric modulation and the drive that controls the mode coupling are weak, so that the system is far from the chaotic regime. Such studies therefore represent a crucial expansion of the canonical framework of parametric resonance [21,37] to encompass the novel effects of controlled nonlinear friction.

In this paper, we realize a parametric oscillator with controlled nonlinear friction using an electromechanical mode with drive-induced resonant nonlinear coupling to another mode, which has a much higher decay rate. A major distinction of such nonlinear friction from the conventional model arises from the dependence of the mode eigenfrequencies on their amplitudes, which is due to conservative mode nonlinearities. Even though the modes are brought to a nonlinear resonance by a drive, the resonance condition is invariably broken as the vibration amplitudes increase. In contrast to the case of conventional nonlinear friction that increases without bound with increasing amplitude, the drive-induced nonlinear friction increases towards a peak value and then decreases. The transition from the nonlinear-friction controlled behavior to the weak nonlinear friction limit qualitatively modifies the whole pattern of parametric resonance, leading to topological changes to the bifurcation diagram. We show, in particular, that the non-conventional nonlinear friction can give rise to two pairs rather than the single pair of coexisting period-two vibration states for conventional nonlinear friction. These states emerge via a swallowtail catastrophe.

## 2. Nonlinear friction controlled via induced mode coupling

### 2.1. Two phonon loss to a faster decaying mode

In the experiment, the micromechanical resonator contains a movable polysilicon plate (100 μm × 100 μm × 3.5 μm) suspended by two beams [figure 1(a)]. The beams have identical width (1.3 μm) and thickness (2 μm) but different lengths. Device A has beams with lengths of 80 μm and 75 μm respectively, while the lengths are 98 μm and 100 μm respectively for device B. In each device, two vibrational modes were used. For mode 1, the plate performs translational motion normal to the substrate as both beams bend [figure 1(b)]. The resonance frequency $\omega_1/2\pi$ and damping constant $\Gamma_1/2\pi$ are ~45 Hz and ~0.5 Hz respectively. For mode 2, the longer beam vibrates parallel to the substrate. It has a significantly higher resonance frequency ($\omega_2/2\pi$ ~ 1 MHz) and a higher damping constant ($\Gamma_2/2\pi \sim 50$ Hz) compared to mode 1. Exact values of $\omega_{1,2}$ and $\Gamma_{1,2}$ are listed in Table 1. The modes are nonlinearly coupled because of the change of the tension in the beam as it deforms under vibrations. In the context of cavity optomechanics [38,39], mode 2 serves as a phonon cavity mode [40,41] that plays a role similar to photon cavity modes.

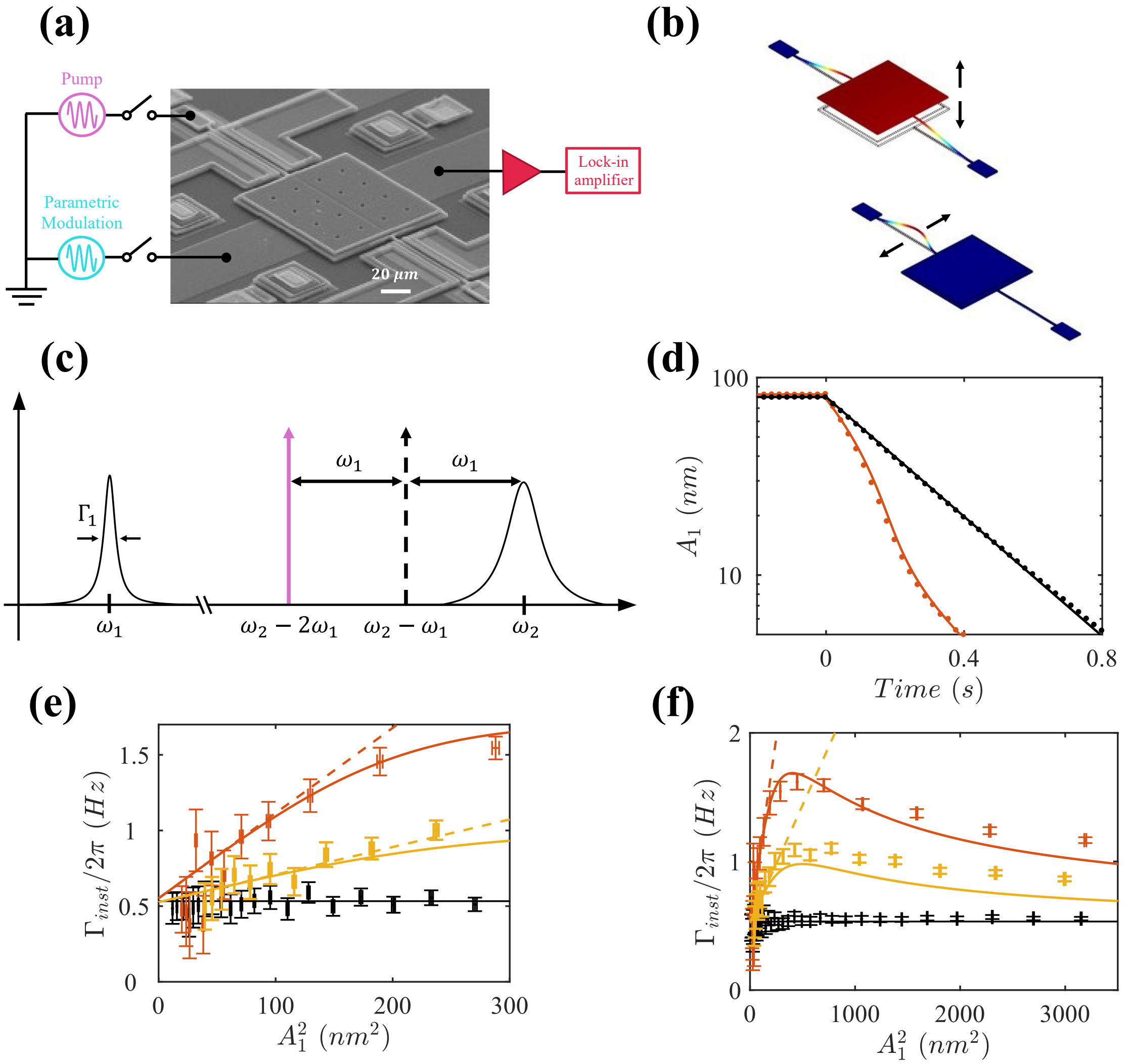

**Figure 1.** Nonlinear friction associated with two phonon loss to a faster decaying mode. (a) Scanning electron micrograph of the electromechanical resonator with a schematic of the excitation and detection circuitry. (b) Mode profiles for mode 1 (top) and mode 2 (bottom). (c) A sketch of the spectra of the two modes. Pumping at $\omega_2 - 2\omega_1$ (purple arrow) transfers two phonons from mode 1 to induce nonlinear friction on it. (d)

Exponential (black) and non-exponential (brown) decay of amplitude $A_1$ with and without sideband pumping in device 1. (e) Changes in the instantaneous decay rate $\Gamma_{\text{inst}} = -d\log|A_1|/dt$ as a function of $A_1^2$ for sideband pump amplitude $F_p$ of 0 (black), 94.9 kN m$^{-2}$ (yellow) and 158.2 kN m$^{-2}$ (brown). (f) Similar plot for larger range of $A_1^2$. Circles are measurement. Dashed straight lines are from equation (6). Solid lines are obtained using the full equations (3) and (4).

| | Short beam | Long beam | $\omega_1/2\pi$ | $\Gamma_1/2\pi$ | $\omega_2/2\pi$ | $\Gamma_2/2\pi$ |
|---|---|---|---|---|---|---|
| Device A | 75 $\mu m$ | 80 $\mu m$ | 43376 Hz | 0.50 Hz | 1580560 Hz | 34.9 Hz |
| Device B | 98 $\mu m$ | 100 $\mu m$ | 47031 Hz | 0.48 Hz | 1130399 Hz | 70.6 Hz |

**Table 1.** System parameters for devices A and B. Device A was initially used until it was mistakenly damaged, after which device B was fabricated to continue the measurement.

Potential differences are applied between the plate and the underlying electrodes. Periodic electrostatic forces on mode 1 can be generated by a small ac voltage at frequency near $\omega_1$ on top of a dc voltage. Alternatively, the spring constant can be parametrically modulated by an ac voltage at frequency near $2\omega_1$ through the spring softening effect. Motion of the plate is detected by measuring the charge flowing into a charge sensitive amplifier as the capacitance between the plate and the electrode changes. It is the parametric resonance of the plate mode that is the central topic of this study. All measurements were performed at pressure of $< 10^{-5}$ torr with the sample chamber submerged in liquid helium.

### 2.2. Tunable conventional nonlinear friction for small amplitudes

We first characterize the dynamics of the coupled modes in the absence of parametric excitation. Nonlinear friction is induced on mode 1 by applying an ac current to the beam [figure 1(a)] to pump mode 2 at the secondary red-detuned sideband [15] at $\omega_F = \omega_2 - 2\omega_1$. In the presence of a perpendicular magnetic field, the Lorentz force modulates the nonlinear coupling between the two modes. The resonant behavior is described by a minimal model with the following equations of motion:

$$\ddot{q}_1 + {\omega_1}^2 q_1 + 2\Gamma_1 \dot{q}_1 + \frac{\gamma_1}{m_1} {q_1}^3 + \frac{\gamma}{m_1} q_1 {q_2}^2 = \frac{F_p}{m_1} 2 q_1 q_2 \, cos(\omega_F t) \quad (1)$$

$$\ddot{q}_2 + {\omega_2}^2 q_2 + 2\Gamma_2 \dot{q}_2 + \frac{\gamma_2}{m_2} {q_2}^3 + \frac{\gamma}{m_2} q_2 {q_1}^2 = \frac{F_p}{m_2} q_1^2 \, cos(\omega_F t) \quad (2)$$

where $q_{1,2}$ denotes the displacements of the plate and the midpoint of the beam, respectively. In mode $i$ ($i$ = 1, 2), $m_i$ represents the effective mass and $\gamma_i$ denotes the Duffing nonlinearity. The last terms on the left-hand side of equation (1) and equation (2) correspond to the dispersive coupling energy $\frac{1}{2}\gamma q_1^2 q_2^2$ of the two modes. $F_p$ is the amplitude of the sideband pumping that comes from the term $-F_p q_1^2 q_2 \, cos(\omega_F t)$ in the Hamiltonian. The system is in the deep resolved-sideband limit, where $\omega_1/\Gamma_2 = 1242$ and 445 for devices A and B respectively.

We convert $q_{1,2}(t), \dot{q}_{1,2}(t)$ into slowly varying complex amplitudes $u_{1,2}(t)$ according to $q_1 - i\omega_1^{-1}\dot{q}_1 = 2u_1 e^{i\omega_1 t}/\sqrt{2m_1\omega_1}$ and $q_2 - i(\omega_F + 2\omega_1)^{-1}\dot{q}_2 = 2u_2 e^{i(\omega_F + 2\omega_1)t}/\sqrt{2m_2\omega_2}$. From equations (1) and (2), the equations for $u_{1,2}$ in the rotating wave approximation read

$$\dot{u}_1 + \Gamma_1 - i\Lambda_{22} u_2 |u_2|^2 - \mathrm{i}\Lambda_{12} u_2 |u_1|^2 = -2i f_p u_1^* u_2 \quad (3)$$

$$\dot{u}_2 + (\Gamma_2 + i\Delta) u_2 - i\Lambda_{22} u_2 |u_2|^2 - i\Lambda_{12} u_2 |u_1|^2 = -i f_p {u_1}^2 \quad (4)$$

where $\Delta = \omega_F - (\omega_2 - 2\omega_1)$, $\Lambda_{11} = \frac{3\gamma_1}{4m_1^2\omega_1^2}$, $\Lambda_{22} = \frac{3\gamma_2}{4m_2^2\omega_2^2}$, $\Lambda_{12} = \frac{\gamma}{2m_1m_2\omega_1\omega_2}$ and $f_p = \frac{F_p}{4m_1\omega_1\sqrt{2m_2\omega_2}}$. The vibration amplitudes in the laboratory frame are given by $A_{1,2} = \sqrt{\frac{2}{m_{1,2}\omega_{1,2}}}\left|u_{1,2}\right|$. $\Delta$ is set to zero for all measurements in this paper.

We start with the case where the vibration amplitude is small so that mode 2 adiabatically follows mode 1. After a short transient, we have $|\dot{u}_2|/|u_2| \ll (\Gamma_2^2 + \Delta^2)^{1/2}$. Disregarding terms $\propto |u_1|^5$ in the equation for $\dot{u}_1$ after $u_2$ has been adiabatically eliminated, we obtain from equations (3) and (4)

$$\dot{u}_1 \approx -u_1(\Gamma_1 + \alpha|u_1|^2) + i\beta u_1|u_1|^2 \qquad (5)$$

The effective decay rate is hence modified to $\Gamma_{\mathrm{ad}} = \Gamma_1 + \alpha|u_1|^2$, where $\alpha = 2f_{\mathrm{p}}^2\Gamma_2 / (\Gamma_2^2 + \Delta^2)$ and $\beta = \Lambda_{11} + 2f_{\mathrm{p}}^2\Delta / (\Gamma_2^2 + \Delta^2)$ [15]. In this limit, conventional nonlinear friction is generated in mode 1. Mode 2 serves as a thermal reservoir to dissipate the energy transferred from mode 1.

**2.3. Non-monotonic nonlinear friction**

We measure the decay rate by recording the ringdown of mode 1. With no sideband pumping ($f_p = 0$), $A_1$ decays exponentially as shown by the black dots in figure 1(d). The instantaneous decay rate $\Gamma_{\mathrm{inst}} = -\mathrm{d}\log A_1/\mathrm{d}t$ is independent of $A_1$, as shown by the black data in figures 1(e) and 1(f). Application of the sideband pump generates nonlinear friction in mode 1. For small amplitudes, the increase in $\Gamma_{\mathrm{inst}}$ is proportional to $A_1^2$, in a manner consistent with conventional nonlinear friction [dash lines in figure 1(e)]. The coefficient for nonlinear friction $\alpha$ is proportional to $f_p^2$ and is therefore readily tunable [yellow and brown data in figure 1(e)].

For larger $A_1$, equation (5) no longer applies. The dependence of $\Gamma_{\text{inst}}$ on $A_1$ becomes non-monotonic. As shown in figure 1(f), $\Gamma_{\text{inst}}$ attains a peak, after which it decreases towards the zero-amplitude value $\Gamma_1$ as $A_1$ further increases. Such non-monotonic dependence arises because the conservative nonlinearities lead to shifts in the eigenfrequencies of the modes comparable to $\Gamma_2$, which in turn disturb the resonant condition required for nonlinear friction [15].

## 3. Response to resonant parametric modulation

### 3.1. Bifurcation diagram for parametric resonance with conventional nonlinear friction

We now study parametric resonance of mode 1, which is excited by modulation of the frequency $\omega_1$, This modulation is implemented using the electrostatic spring softening effect, with an ac voltage at frequency $2\omega_h$ near $2\omega_1$ applied to one of the electrodes underneath the top plate. The effect of the modulation can be described by modifying the term ${\omega_1}^2 q_1$ in equation (1),

$$\omega_1^2 q_1 \rightarrow \omega_1^2[1 + h\cos(2\omega_h t)]q_1, \quad |\omega_1 - \omega_h| \ll \omega_1.$$

Here $h$ characterizes the modulation amplitude. We denote the detuning of the parametric modulation frequency by $\epsilon = \omega_h - \omega_1$. When the friction is linear, sufficiently strong parametric modulation excites vibrations at $\omega_h$. The excitation occurs via a supercritical pitchfork bifurcation $b_1$ in figure 2(a), where $\epsilon = \epsilon_{b_1}$. At this frequency detuning, a pair of stable period-two vibration states emerge, while the zero-amplitude state becomes unstable. At the subcritical pitchfork bifurcation $b_2$ ($\epsilon = \epsilon_{b_2}$), the unstable zero-amplitude state becomes stable again, with the emergence of a pair of unstable period-two vibrational states.

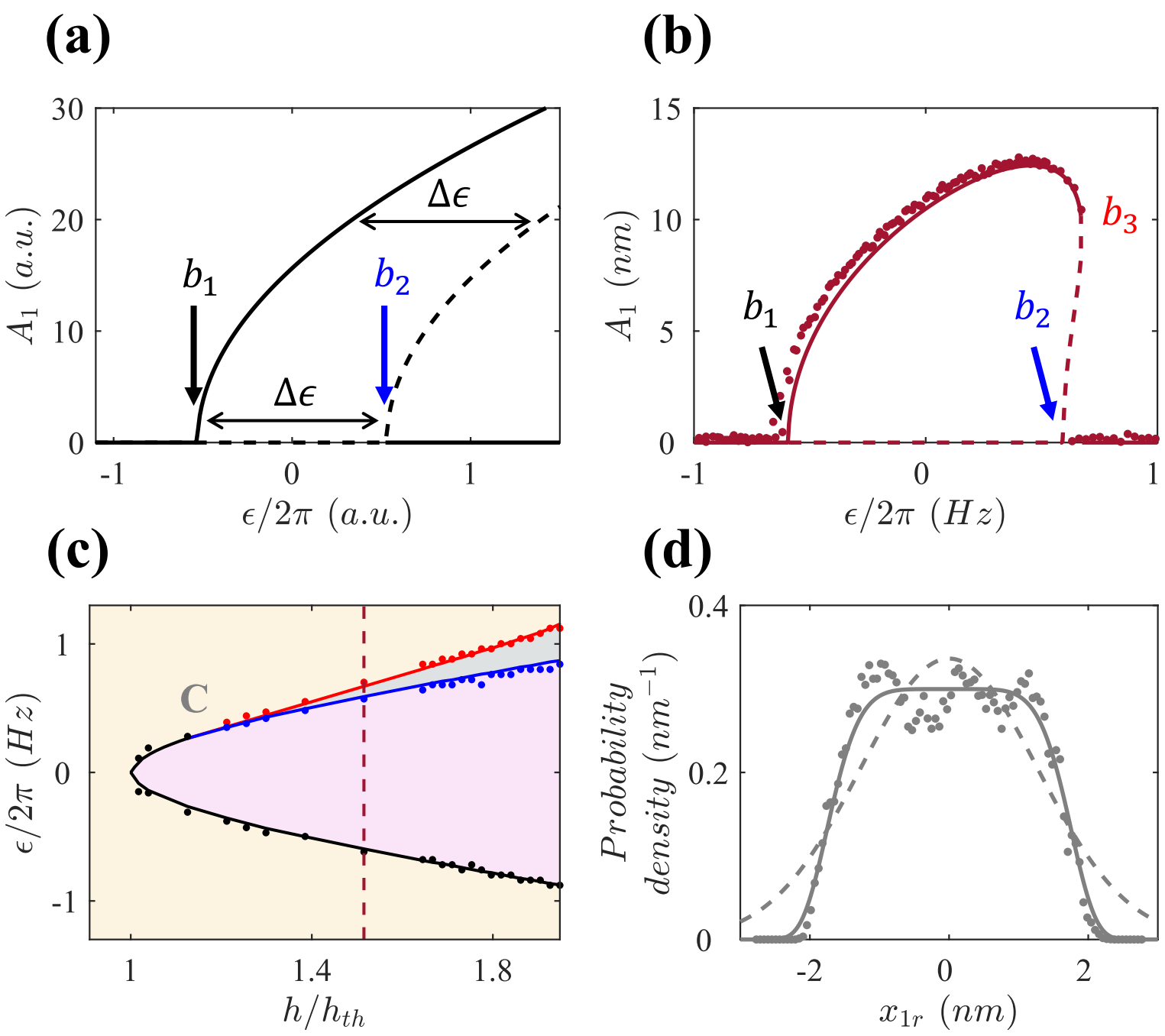

**Figure 2.** Parametric resonance with conventional nonlinear friction. (a) Amplitude of the parametrically excited state vs. detuning frequency $\epsilon$ of parametric modulation for stable (solid) and unstable (dashed) period-two vibrations for parametric resonance with linear friction. We use $b_n$ to indicate the $n$th bifurcation point. $b_1$ and $b_2$ indicate the supercritical and subcritical pitchfork bifurcations respectively. $\Delta\epsilon$ is constant at different amplitude for fixed $\Gamma_1$ (horizontal arrows). (b) Similar plot for device 1 in the regime of conventional nonlinear friction, for the modulation amplitude $h/h_{th}$ = 1.52, where $h_{th} = 4\Gamma_1/\omega_1$ is the threshold for the onset of period-two vibrations. The side-band drive amplitude $F_p$ is 94.9 kN m$^{-2}$. No vibration states exist beyond bifurcation point $b_3$. (c) Bifurcation diagram with parameters $\epsilon$ and $h$. Solid lines represent $b_1$ (black), $b_2$ (blue) and $b_3$ (red) from theory. Circles are measurement. The colors yellow, pink and grey correspond to parameter space with a stable zero-amplitude state, a pair of stable period-two states, and coexistence of the stable zero-amplitude state and the pair of stable period-two states, respectively. The amplitude response in (b) is obtained along the vertical dashed line. $b_2$ and $b_3$ merge at the critical point C. (d) At C, fluctuations are strongly non-Gaussian. Dashed and solid lines are quadratic and 6$^{th}$ order fits to the logarithm of the probability distribution.

In the linear-friction limit, both the stable and unstable branches of the response extend to infinitely large vibration amplitudes. The separation in frequency $\Delta\epsilon$ between the stable and unstable vibrational states $\Delta\epsilon = 2\sqrt{(\omega_1 h/4)^2 - \Gamma_1^2}$ is independent of the vibration amplitude $A_1$, as shown in figure 2(a). Vibrations can only be excited if the damping is smaller than $\omega_1 h/4$.

Next, we turn on the sideband pumping to study the effects of nonlinear friction on parametric resonance. A simple way to understand the effects of conventional nonlinear friction is to consider it as the increase in the effective damping with amplitude. In this picture, as $A_1$ increases, $\Delta\epsilon$, as given by the above expression with $\Gamma_1$ replaced with effective damping, decreases. When $\Delta\epsilon$ reaches zero, each of the two stable period-two vibration states merges with the corresponding unstable state in a saddle-node bifurcation. In figure 2(b), this saddle-node bifurcation is labelled as $b_3$, at frequency detuning $\epsilon_{b_3}$. For $\epsilon > \epsilon_{b_3}$, only the zero-amplitude state is stable for conventional nonlinear friction. The sharp jump of vibration amplitude to zero at $b_3$ has also been measured in other parametric oscillators with nonlinear friction [6,42,43].

Nonlinear effects are negligible at the subcritical pitchfork bifurcation $b_2$. As $\epsilon$ is swept up and down between $\epsilon_{b_2}$ and $\epsilon_{b_3}$, there is hysteresis in the vibration amplitude. For smaller values of parametric drive $h$, the hysteresis region shrinks as $\epsilon_{b_2}$ and $\epsilon_{b_3}$ approach each other. Figure 2(c) shows the bifurcation diagram where $\epsilon_{b_2}$ and $\epsilon_{b_3}$ are denoted in blue and red respectively. Squares are measured values and solid lines are calculated using equations (1) and (2). At the critical point C, $\epsilon_{b_2}$ and $\epsilon_{b_3}$ merge. Here, two stable period-two states, two unstable period-two states and the zero-amplitude state merge together. The point C is located where all 3 regions on the bifurcation diagram meet. We directly measure the merging $\epsilon_{b_2}$ and $\epsilon_{b_3}$ in our experiment by tuning both the amplitude and frequency detuning of the parametric modulation to the critical values of $\epsilon_c$ and $h_c$ respectively.

Near the critical point the effective potential becomes extremely shallow [44]. Figure 2(d) presents numerical calculations of the amplitude fluctuations at the critical point. We find that the stationary probability distribution becomes anomalously broad and strongly non-Gaussian. Furthermore, the top of the logarithm of the distribution no longer behaves as an inverted parabola that is typical for Gaussian distributions. Instead, a power law of 6th order is required to fit it. The approaching of 5 stationary states as the parameters approach $(h_c, \epsilon_c)$ is shown in Supplementary figure 3.

**3.2. Isolated branch in parametric resonance**

An important characteristic of the nonlinear friction in our system, and presumably in all systems where nonlinear friction is achieved by induced mode coupling, is that the instantaneous decay rate $\Gamma_{\mathrm{inst}}$ does not increase monotonically. This leads to a qualitative change of the response to parametric excitation compared to the conventional picture. As the vibration amplitude is further increased, $\Gamma_{\mathrm{inst}}$ attains a maximum and then reverts to the zero-amplitude value [figure 1(e)]. Consequently, it is possible for the vibrations that became unstable for $\epsilon > \epsilon_{b_3}$ to become stable again at large amplitudes.

Moreover, we find that a pair of period-two stable states can appear in this range with the varying detuning $\epsilon$. The states have equal nonzero amplitudes and opposite phases. They emerge along with a pair of unstable period-two states via saddle-node bifurcations, distinct from the conventional pitchfork bifurcation for the onset of period-two states of a parametric oscillator. The corresponding saddle-node bifurcation is marked by $b_4$ in figure 3(a). Such isolated branches in the response are known as isola [45,46]. They are isolated from the main response and cannot be reached by sweeping $\epsilon$ alone.

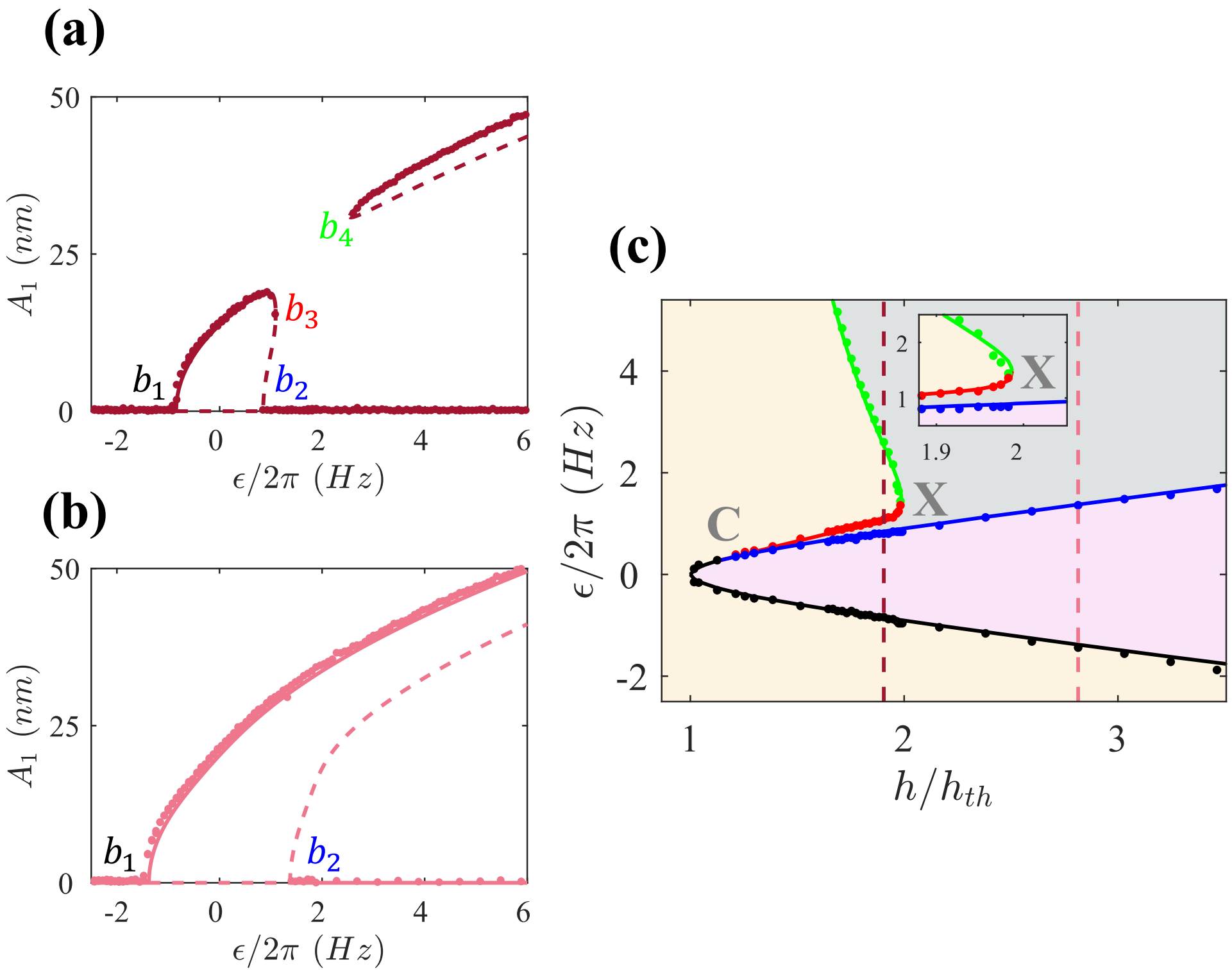

**Figure 3.** Isolated branch of period-two states for non-monotonic nonlinear friction. (a) Amplitude vs. $\epsilon$ for induced nonlinear friction in device 1, for $h/h_{th}$ = 1.90 and $F_p$ = 94.9 kN m$^{-2}$. The isolated branch starts at bifurcation point $b_4$. As $h$ is increased, $b_3$ and $b_4$ move towards each other and merge. (b) At $h/h_{th}$ = 2.81, the frequency response qualitatively resembles that of a parametric oscillator without nonlinear friction. (c) In the bifurcation diagram, $b_3$ and $b_4$ (green) merge at point X. The amplitude responses in (a) and (b) are obtained along the dark red and light red vertical dash lines respectively. Inset: close up of the region where $b_3$ and $b_4$ merge.

If $h$ is increased to reach a threshold value, the energy supplied by the parametric drive can eventually compensate for the energy loss at all amplitudes. The separation between the isola and the main response shrinks to zero as saddle-node bifurcations $b_3$ and $b_4$ merge together. Upon further increase of $h$, the response shown in figure 3(b) qualitatively resembles that of a parametric oscillator under linear friction. In the bifurcation diagram figure 3(c), the extra bifurcation $b_4$ is represented as the green branch that comes in from the top to merge with bifurcation $b_3$ (red). Such behavior of the bifurcation diagram was measured in both devices 1 and 2.

### 3.3. Swallowtail catastrophe

At where the isola branches approach the conventional-nonlinear-friction branches, i.e., where the bifurcation points $b_3$ and $b_4$ are close, the system is very sensitive to the strength of the mode coupling. This strength is controlled by the sideband pump. We now show that, by varying the secondary sideband pump amplitude $f_p$, the nonlinear friction can be modified to reveal extra branches of period two states. In other words, the system can have not just one pair of period-two states, but two pair of such states. The onset of such states is associated with a swallowtail catastrophe [47] that we observe in both devices 1 and 2.

The emergence of the extra states can be understood from the bifurcation diagram shown in figures 4(a) and 4(b), which refer to device 2. These diagrams have the form that is characteristic of the swallowtail. The major difference from figure 3(b) is that a light-blue triangular region has emerged at the intersection of the red and green bifurcation curves [figure 4(b)]. It is in this region that mode 1 displays an extra set of stable period-two vibrations, at an amplitude different from the first set of period-two states. There are nine coexisting stationary points: two pairs of stable period-two vibration states, two pairs of unstable period-two vibration states and one stable zero-amplitude state. Since the states in a period-two pair have the same amplitudes and opposite phases, it is sufficient to follow one state of a pair.

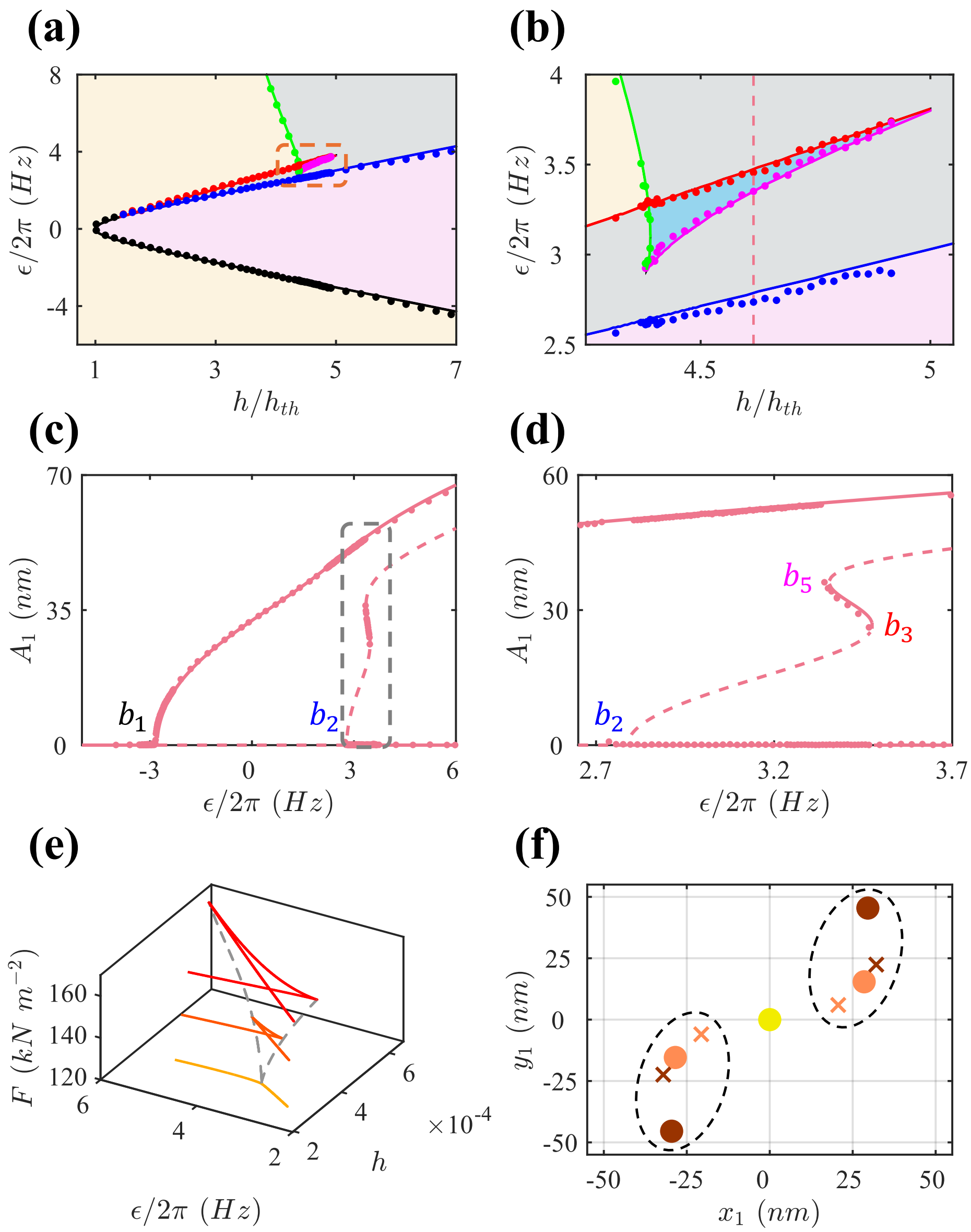

**Figure 4.** Emergence of a second pair of stable states of parametric resonance in a swallowtail catastrophe. (a) The bifurcation diagram for device 2 at $F_p$ = 145 kN m$^{-2}$ displays a specific structure near the crossing point of the red and green bifurcation lines. Lines and circles represent theory and measurement respectively. (b) Close up of the dashed box in (a). In the light blue region the system has two, rather than the conventional one pairs of period-two states. They coexist with the zero-amplitude state. A purple bifurcation line, which was not present in figure 3, has emerged. (c) Amplitude vs. $\epsilon$ for $h/h_{th}$ = 4.61. An extra pair of period-two states emerges between bifurcations $b_5$ and $b_3$. Measurement is done along the vertical dash line in (b). (d) Close-up near $b_5$ and $b_3$: measured data is shown by dots, solid line is the theory. (e) Merging of two

bifurcation lines (solid) in the $\epsilon$-$h$ plane gives a cusp point. When a third parameter $f_p$ of the electromechanical oscillator is varied, the cusp points move along the dashed lines. Swallowtail catastrophe occurs when two cusp points merge together. (f) Pairs of stable period-two states are represented by the full circles in the $(x_1 y_1)$ phase plane of mode 1 in the rotating frame. The unstable states are represented by crosses. Swallowtail catastrophe involves merging of two stable states and two unstable states surrounded by a dashed boundary.

We illustrate the coexisting states in figures 4(c) and 4(d) by choosing $h/h_{th} = 4.61$ near the middle of the triangular region, indicated by the vertical dash line in figure 4(b). Starting from zero amplitude, the unstable branch [represented by the lower dashed line in figures 4(c) and 4(d)] initially bends on the $(\epsilon, A_1)$ −plane towards high frequencies as usual when the vibration amplitude $A_1$ increases. At intermediate amplitudes of ~ 6 nm, bifurcation point $b_3$, there occurs a saddle-node bifurcation where the unstable state on this branch merges with a stable state. The amplitude of the stable state increases with the *decrease* of $\epsilon$. Once $(\epsilon, A_1)$ reach the bifurcation point $b_5$, there occurs another saddle-node bifurcation and, on the emerging unstable branch, $A_1$ again increases with $\epsilon$. The evolution of the pattern of stationary states on the phase plane with the increasing $\epsilon$ is shown in Supplementary figure 3.

We find that if the sideband pump amplitude $f_p$ is reduced, the light blue region in the extra pattern of the bifurcation diagram shrinks and ultimately disappears in a swallowtail catastrophe. Figure 4(e) shows the evolution of the bifurcation lines, while Figure 4(f) shows the merging of 4 pairs of stationary states on the phase plane as the swallowtail catastrophe is approached. In figure 4(e) the solid lines represent the bifurcation lines where two stationary states merge for a fixed $h$. The dashed lines show locations of the cusps where two saddle-node bifurcations merge as $h$ is varied. The swallowtail catastrophe corresponds to merging of the two cusps. In our system, the larger-$h$ cusp point is formed by the merging of red and purple bifurcation lines at the top right-hand corner of the light-blue triangular region in the bifurcation diagram [Figure 4(b)]. The smaller-$h$ cusp point involves the merging of the green and purple bifurcation lines at the lower left-hand corner of the triangular region. While the system can be placed at a cusp point by choosing $\epsilon$ and $h$ of the parametric drive, bringing the two cusp points together to achieve a swallowtail catastrophe requires sweeping the third parameter $f_p$.

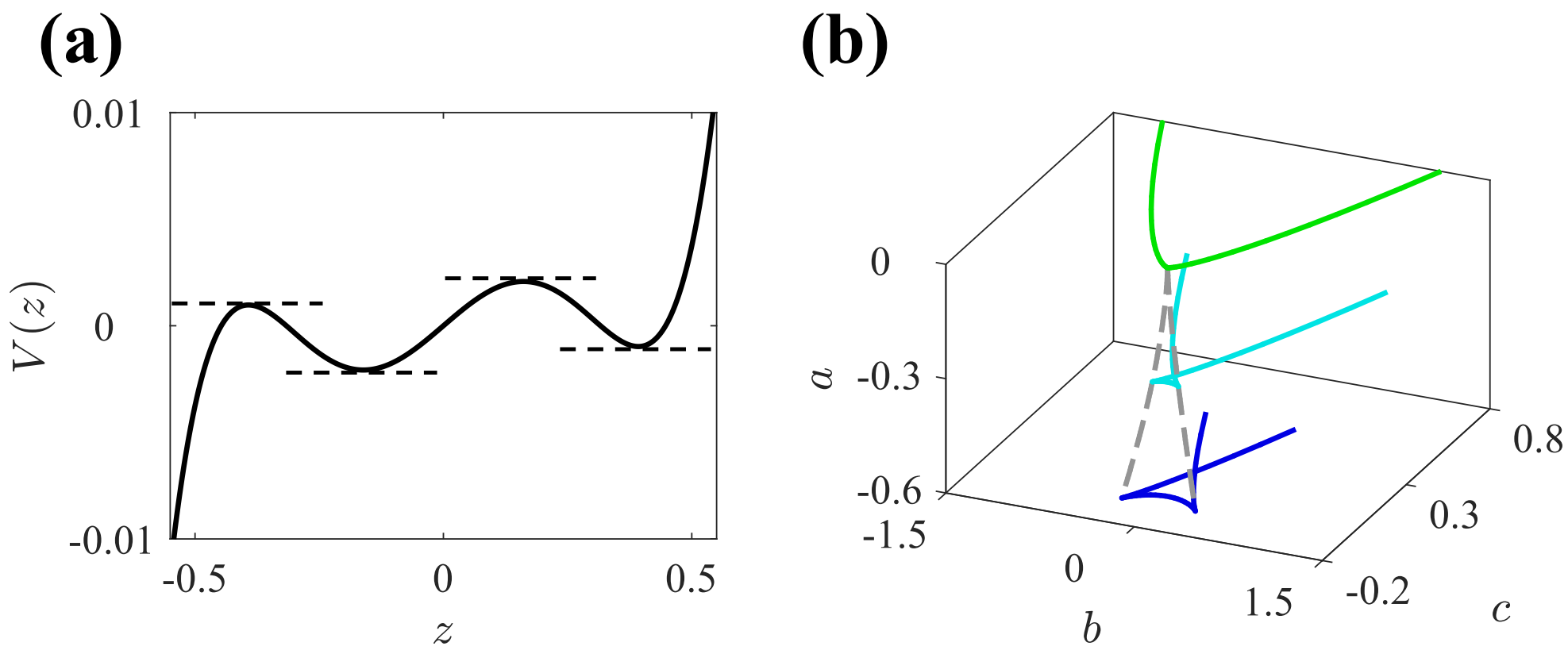

**Figure 5.** Normal form potential of swallowtail catastrophe. (a) Coexistence of two maxima and two minima for the normal form potential $V_{eff}(z)$ of swallowtail catastrophe, with $a$ = -0.3, $b$ = 0 and $c$ = 0.02. (b) Solid lines are bifurcation lines in the $b$-$c$ plane for $a$ = 0.1 (green), -0.1 (light blue) and -0.3 (blue). Dash lines indicate cusp ridges where two bifurcation points meet. The two cusp ridges merge at the swallowtail catastrophe.

The potential for swallowtail catastrophes has the normal form $V(x;a,b,c) = x^5 + ax^3 + bx^2 + cx$ with three control parameters $a$, $b$ and $c$ [47]. These parameters are linear combinations of the control parameters of the system, the amplitude $h$ and frequency detuning $\epsilon$ of the modulating field and the sideband pump amplitude $f_p$. The potential $V(x; a, b, c)$ and the evolution of the bifurcation lines in the $(b, c)$-plane is shown in figure 5.

## 4. Discussion

The results reported here are of interest from two complementary perspectives. First, they demonstrate the coexistence of two distinct pairs of period-two vibrational states in a parametrically driven oscillator. This phenomenon arises from nonlinear friction induced by a widely used implementation in which the oscillator is coupled to a more strongly damped higher-frequency auxiliary mode through a drive tuned near the difference between the eigenfrequency of the auxiliary mode and twice the oscillator eigenfrequency. The emergence of multiple period-two states is rooted in the interplay between the frequency dependence of the effective coupling strength and the amplitude-dependent frequency shifts produced by the intrinsic nonlinearities of the oscillator and the mode.

Second, the multiplicity comes along with the nontrivial topology of the bifurcation diagram of the oscillator and provides a means to explore this topology in a controlled fashion. We show that the multiplicity emerges via a swallowtail catastrophe, revealing new manifestations of catastrophe theory -- a mathematical framework that describes abrupt, discontinuous transitions in complex systems. The swallowtail catastrophe has been invoked to explain phenomena across diverse fields, including psychology [48], ecology [49] and geology [50]. However, experimental observation of the swallowtail catastrophe is challenging because it requires precise control of three independent parameters to map the system response. As a result, direct measurements have been reported in only a few physical systems, including optical beams [51,52] and non-reciprocal electronic circuits [53]. Our results establish electromechanical oscillators as a versatile platform for revealing peculiar, sometimes non-obvious consequences of the bifurcation topology and quantitatively characterizing the underlying dynamics.

Bringing a system close to swallowtail catastrophe opens a way of exploring new critical fluctuation phenomena and revealing new scaling behavior of fluctuations in nonlinear systems, and in particular in systems far from thermal equilibrium. For several types of bifurcations, where two or maximum three states come close to each other in phase space, fluctuation scaling has been thoroughly studied using nonequilibrium systems ranging from driven mechanical oscillators [54-58] and particles in traps [59,60] to modulated superconducting circuits [61]. In contrast, near swallowtail catastrophe, four stationary states come close to each other in phase space (in our system, there are two sets of four close states). A rich pattern of fluctuations ensues. For example, external noise can lead to switching both between close stable states and globally away from these states to the zero amplitude states. Our system provides a platform for studying how the switching rates scale with the three controllable bifurcation parameters.

Our findings open new opportunities for exploiting controlled nonlinear friction in parametrically driven oscillators. In particular, they enable the realization of nontrivial bifurcation topologies, including a swallowtail catastrophe, and provide a platform for investigating the associated dynamics and fluctuation phenomena far from thermal equilibrium. A key result is the emergence of a new form of parametric resonance, in which two pairs of period-two states coexist, extending the canonical picture of parametric excitation. The ability to engineer and control multiple coexisting vibrational states could be leveraged in applications of parametric oscillators ranging from sensing to information processing, including multistate information encoding.

**Acknowledgements**

This work is supported by the Research Grants Council of Hong Kong SAR, China (Project No. 16301421). M.I.D. acknowledges partial support from the Gordon and Betty Moore Foundation (Award No. GBMF12214). We thank S. W. Shaw for useful discussions.

**Data availability statement**

All data that support the findings of this study are included within the article.

**Appendix A. Parametric modulation**

Voltage $V = V_{dc} + V_{ac} \cos(\omega_h t)$ applied between the plate and the electrode on the right-hand side generates electrostatic force $F = \frac{1}{2}\frac{dC}{dq_1}V^2$, where $C$ is the capacitance between the plate and the electrode. The dc voltage $V_{dc}$ leads to a static displacement $q_{1,0}$ of the plate. Expanding the force about $q_{1,0}$ gives $F = \frac{1}{2}(C' + C''q_1)V^2$ where $q_1$ is the displacement measured from $q_{1,0}$, $C'$ and $C''$denote the derivatives of $C$ with respect to $q_1$. Higher order terms are neglected. The parametric modulation at frequency $2\omega_h$ originates from the second term in $F$, with $h = \frac{1}{m_1\omega_1^2} C''V_{dc}V_{ac}$.

**Appendix B. Amplitudes of the period-two states**

In the presence of parametric modulation, the term ${\omega_1}^2 q_1$ in equation (1) is modified to ${\omega_1}^2[1+hcos(2\omega_h t)]q_1$. For sufficiently large $h$, there are excited oscillations at $\omega_h$. Theoretical analysis is performed by converting $q_{1,2}(t), \dot{q}_{1,2}(t)$ into slow complex amplitudes $v_{1,2}(t)$ defined as $v_1(t) = \sqrt{\frac{m_1}{2\omega_1}}(\omega_1 q_1 - i\dot{q}_1)\exp(-i\omega_h t)$ and $v_2(t) = \sqrt{\frac{m_2}{2(\omega_F+2\omega_h)}}[(\omega_F + 2\omega_h)q_2 - i\dot{q}_2]\exp[-i(\omega_F + 2\omega_h)t]$. Substituting into equations (1) and (2) and applying the rotating wave approximation to ignore fast rotating terms gives the equations of motion for $v_{1,2}$:

$$\dot{v}_1 + (\Gamma_1 + i\epsilon)v_1 - i\Lambda_{11}|v_1|^2 v_1 - i\Lambda_{12}|v_2|^2 v_1 = -2if_p v_1^* v_2 + ih_d v_1^* \quad \text{(B1)}$$

$$\dot{v}_2 + [\Gamma_2 + i(\Delta + 2\epsilon)]v_2 - i\Lambda_{22}v_2|v_2|^2 - \mathrm{i}\Lambda_{12}v_2|v_1|^2 = -if_p {v_1}^2 \quad \text{(B2)}$$

where $h_d = \frac{h\omega_1}{4}$. The stationary states are obtained by setting $\dot{v}_1 = \dot{v}_2 = 0$. From equations (B1) and (B2), it can be seen that if $(v_1, v_2)$ is a solution, $(-v_1, v_2)$ is also a solution. The amplitudes of the stationary states are given by the solutions of the following algebraic equations:

$$\begin{cases} (\epsilon|v_1|^2 - 4\epsilon|v_2|^2 - 2\Delta|v_2|^2 - \Lambda_{11}|v_1|^4 + \Lambda_{12}|v_1|^2|v_2|^2 + 2\Lambda_{22}|v_2|^4)^2 + (\Gamma_1|v_1|^2 + 2\Gamma_2|v_2|^2)^2 = h_d^2|v_1|^4 \\ [\Gamma_2^2 + (\Delta + 2\epsilon - \Lambda_{22}|v_2|^2 - \Lambda_{12}|v_1|^2)^2]|v_2|^2 = f_p^2|v_1|^4 \end{cases} \quad (B3)$$

The stability of the stationary states is determined by linearizing equations (6) and (7) about the stationary values of $v_{1,2}$.

**Appendix C. Reducing the system to the normal form of swallowtail catastrophe**

Swallowtail catastrophe has codimension 3. For the swallowtail catastrophe in our experiment, the three control parameters are $\epsilon, f_p$ and $h_d$. In figure 4(e) the solid lines represent the bifurcation lines where two stationary states merge for a fixed $h_d$ and the dashed lines show locations of the cusps where two saddle-node bifurcations merge as $h_d$ is varied. We determine the parameters at which swallowtail catastrophe occurs by locating the intersection of the two dashed lines in figure 4(e). The corresponding complex amplitudes for the stationary state are $v_j = v_j^{sw}, j = 1, 2$.

Reduction of the equations into the normal form of the swallowtail catastrophe requires linearizing the equations of motion about $v_j^{sw}$ to identify the soft mode. We write the complex amplitudes as $v_j = x_j + \mathrm{i}\, y_j$ where $x_j$ and $y_j$ are real. The equations of motion can be rearranged to take the form:

$$\dot{x}_1 = F_1(x_1, y_1, x_2, y_2, \vec{a})$$

$$\dot{y}_1 = F_2(x_1, y_1, x_2, y_2, \vec{a})$$

$$\dot{x}_2 = F_3(x_1, y_1, x_2, y_2, \vec{a})$$

$$\dot{y}_1 = F_4(x_1, y_1, x_2, y_2, \vec{a}) \tag{C1}$$

where the components of the vector $\vec{a}$ are $a_1 = \epsilon, a_2 = f_p, a_3 = h_d$. We linearize the equations about the stationary point $v_j^{sw}$. The new variables $w_1 = x_1 - x_1^{sw}, w_2 = y - y_1^{sw}, w_3 = x_2 - x_2^{sw}, w_4 = y_2 - y_2^{sw}$ follows the equation of motion:

$$\dot{w}_n = P_n(\boldsymbol{w}, \vec{a}) \tag{C2}$$

where vector $\boldsymbol{w}$ has components $w_1, w_2, w_3, w_4$ and $P_n$ is determined by functions $F_n$. At the swallowtail catastrophe $\vec{a} = \vec{a}^{sw}$, one of the eigenvalues of the $4 \times 4$ matrix of $\left(\frac{\partial P_n}{\partial w_m}\right)$ is zero. The corresponding eigenvector is the slow mode, it is a linear combination of the components $w_1, \dots, w_4$, $z = \sum c_i w_i$ .

After rescaling, the equation of motion for $z$ is $\dot{z} = -\frac{\partial V}{\partial z}$. The effective potential $V_{eff}$ is given by:

$$V_{eff} = \frac{1}{5}z^5 + \frac{1}{3}\,az^3 + \frac{1}{2}\,bz^2 + cz \qquad \text{(C3)}$$

The coefficients $a, b, c$ are linear combinations of $a_i - a_i^{sw}$, and thus are linear combinations of the deviations of the three control parameters $\epsilon, f_p$ and $h_d$ from their values at the swallowtail catastrophe.

Figure 5(a) shows $V(z)$ for $a$ = -0.3, $b$ = 0 and $c$ = 0.02. In figure 5(b), the solid lines represent the bifurcation points where two equilibrium points coalesce, for three different values of $a$. Each of the light and dark blue curves possesses two cusp points where two bifurcation point merges at the two corners of the triangle. The dashed lines are cusp ridges that indicate the location of the cusp points as parameter $a$ is varied. Swallowtail catastrophe takes place at $a$ = 0 when the two cusp ridges come together. All four equilibrium points merge together at the swallowtail catastrophe.

**Appendix D. Fluctuations at the critical point**

Calculations of the fluctuations at the critical point [figure 2(d)] are performed by adding random noise $\xi_{1,2}(t)/m_{1,2}$ to the right-hand side of equations (1) and (2) respectively, where $\langle \xi_i(t)\xi_j(t')\rangle = 4m_i k_B T \delta_{ij}\delta(t - t')$ and T = 50 K.

**Appendix E. Exciting the resonator to the isolated branches**

The dark red curve in figure 3(a) contains an upper branch that is isolated from the main peak. It is not possible to excite the system to the isolated branch by merely sweeping the frequency of the parametric modulation $\varepsilon$ while keeping the modulation amplitude fixed.

Instead, it is necessary to first excite the system to large vibration amplitudes and then reduce the modulation amplitude. One way to access the isolated branch is to first sweep $\varepsilon$ at a larger modulation amplitude where the bifurcations $b_3$ and $b_4$ have merged, and then decrease the modulation amplitude to the target value. For example, one can choose $h/h_{th}$ =2.81 and sweep $\varepsilon$ up along the light red curve in figure 3(b) to reach $A_1$ ~ 22 nm at $\epsilon/2\pi$ = 4 Hz. The modulation amplitude can then be slowly reduced while keeping ε constant to bring the system to the brown isolated branch of the dark red curve at $h/h_{th}$ = 1.90 in figure 3(a).

The procedure to access the isolated branch between $b_5$ and $b_3$ is discussed in Supplementary Note 2.

**Data availability**

All the data supporting the findings of this study are available within the article and its Supplementary Information or from the corresponding author upon reasonable request.

# Supplementary Material:

## Multibranched parametric resonance and swallowtail catastrophe in electromechanical oscillators with nonlinear friction

Chan *et al.*

Supplementary Note 1. Emergence of bifurcation $b_6$

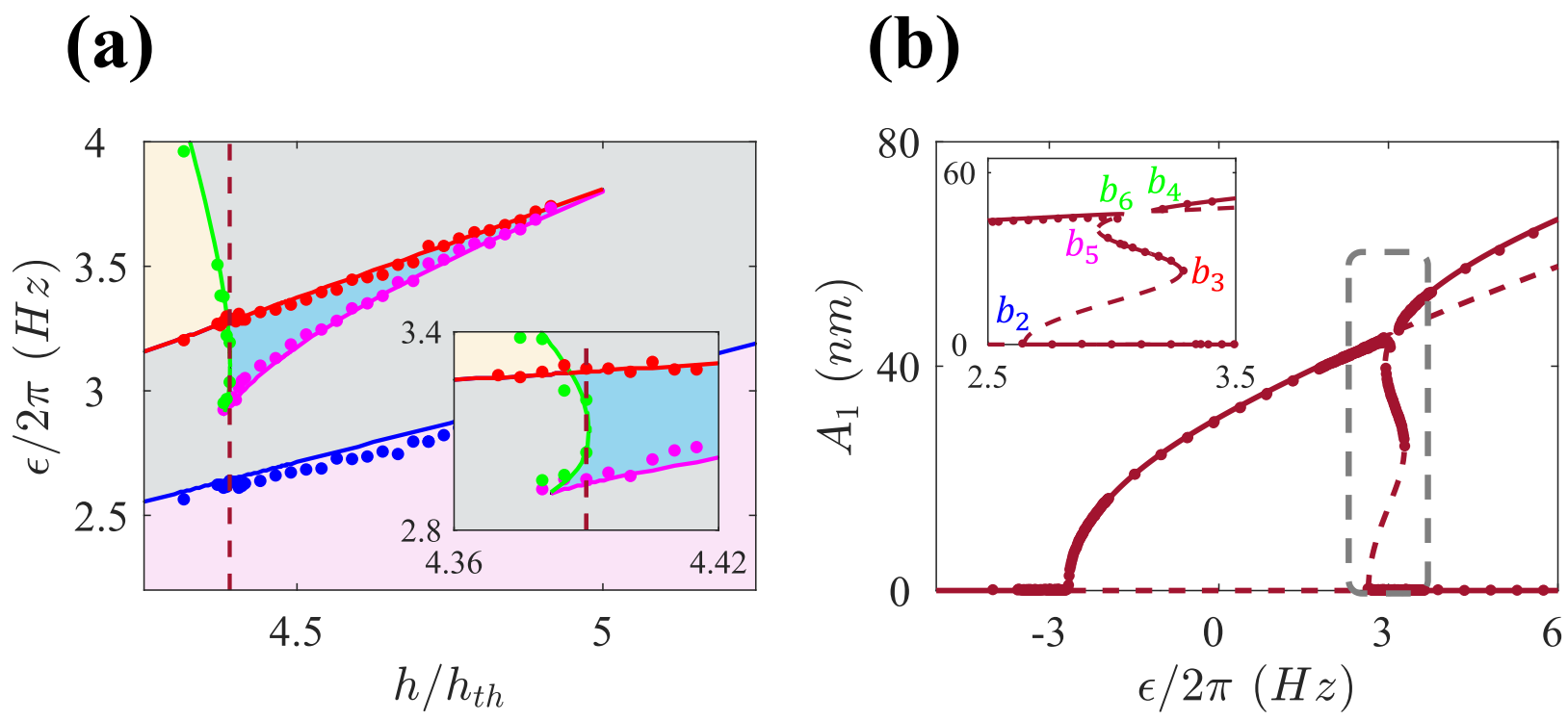

**Supplementary Figure 1.** Different branches of responses near a swallowtail catastrophe. (a) The bifurcation diagram near the extremum of the green bifurcation line as function of $h$, for device 2 at $F_p$ = 145 kN m$^{-2}$. In the light blue region two pairs of period-two states and the zero-amplitude state coexist. Lines and circles represent theory and measurement respectively. Inset: Close-up to show the merging of the green and purple bifurcation lines at a cusp point. (b) Amplitude vs. $\epsilon$ for the modulation amplitude $h$ smaller then its value at the extremum of the green bifurcation line, $h/h_{th} = 4.39$. Measurement is done along the vertical dash line in (a). Bifurcation point $b_6$ appears. Inset: Close-up to show new bifurcation point $b_6$.

At the swallowtail catastrophe, an second pair of period-two stable states emerges. The dependence of the amplitude on parametric drive detuning $\epsilon$ is shown in figure 4c for $h/h_{th} = 4.61$. At slightly smaller $h/h_{th}$ of 4.39 [Supplementary figure 1(b)], the top branch in the response separates into two segments, in a fashion similar to figure 3(a). On the right-hand side, the segment corresponds to the isolated branch at high amplitudes discussed earlier that ends at the low frequency side with saddle-node bifurcation $b_4$. The segment on the left-hand side connects to the main parametric response at low amplitudes. As $\epsilon$ increases, the stable state merges with an unstable state at bifurcation $b_6$.

As $h$ increases, $b_5$ and $b_6$ merge in a co-dimension 2 bifurcation.

Supplementary Note 2. Exciting the resonator to the second branch of coexisting pairs of period-two states

Accessing the isolated branch between $b_5$ and $b_3$ in figures 4(c) and 4(d) requires setting $h/h_{th}$ to the smaller value of 4.39 first so that the upper branch is terminated at bifurcation $b_6$ similar to Supplementary figure 1(b). When $\epsilon$ is swept up past $b_6$, the system jumps to the middle branch. $h/h_{th}$ is then gradually changed back to 4.61 to record the data in figures 4(c) and 4(d).

Supplementary Note 3. Merging of five states at critical point C

With ordinary nonlinear friction, the pair of period-two stable states, the pair of period-two unstable states and the zero amplitude states all merge together at critical point C. Since five states merge together, fluctuations at this point are particularly strong. The logarithm of the stationary distribution scales as $x_{1r}^6$, see figure 2(d) of the main text. Here $x_{1r}$ is the linear combination of the quadratures of the oscillator that describes the soft mode in the rotating frame.

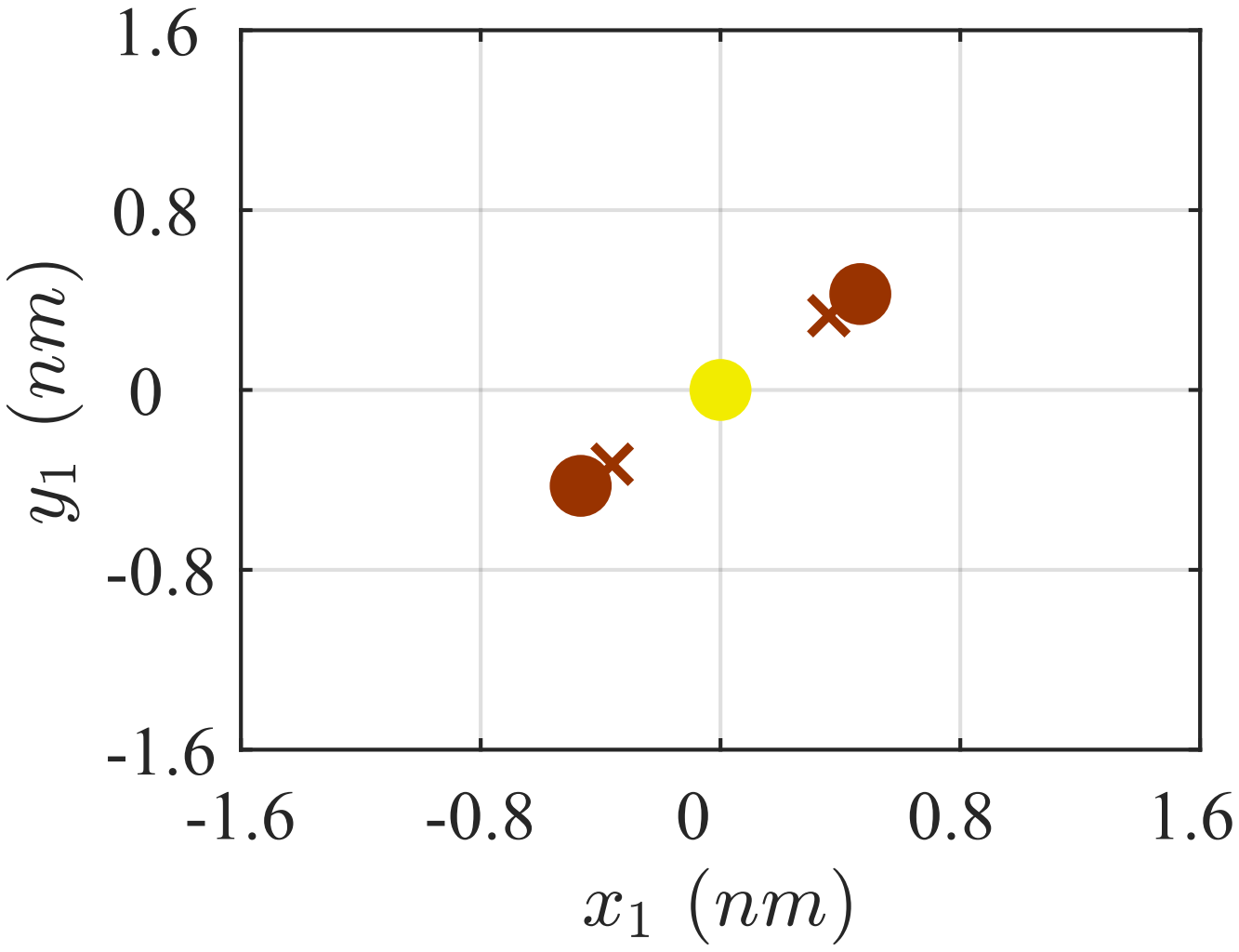

**Supplementary Figure 2.** Merging of 5 states at critical point C.

Supplementary Note 4. Emergence of second pair of period-two states on the phase plane.

In figure 4(d), the second pair of stable period-two states exists between parametric drive frequency detuning $\epsilon_{b5}$ and $\epsilon_{b3}$. Supplementary figure 3 shows the emergence of second pair of period-two states on the phase plane.

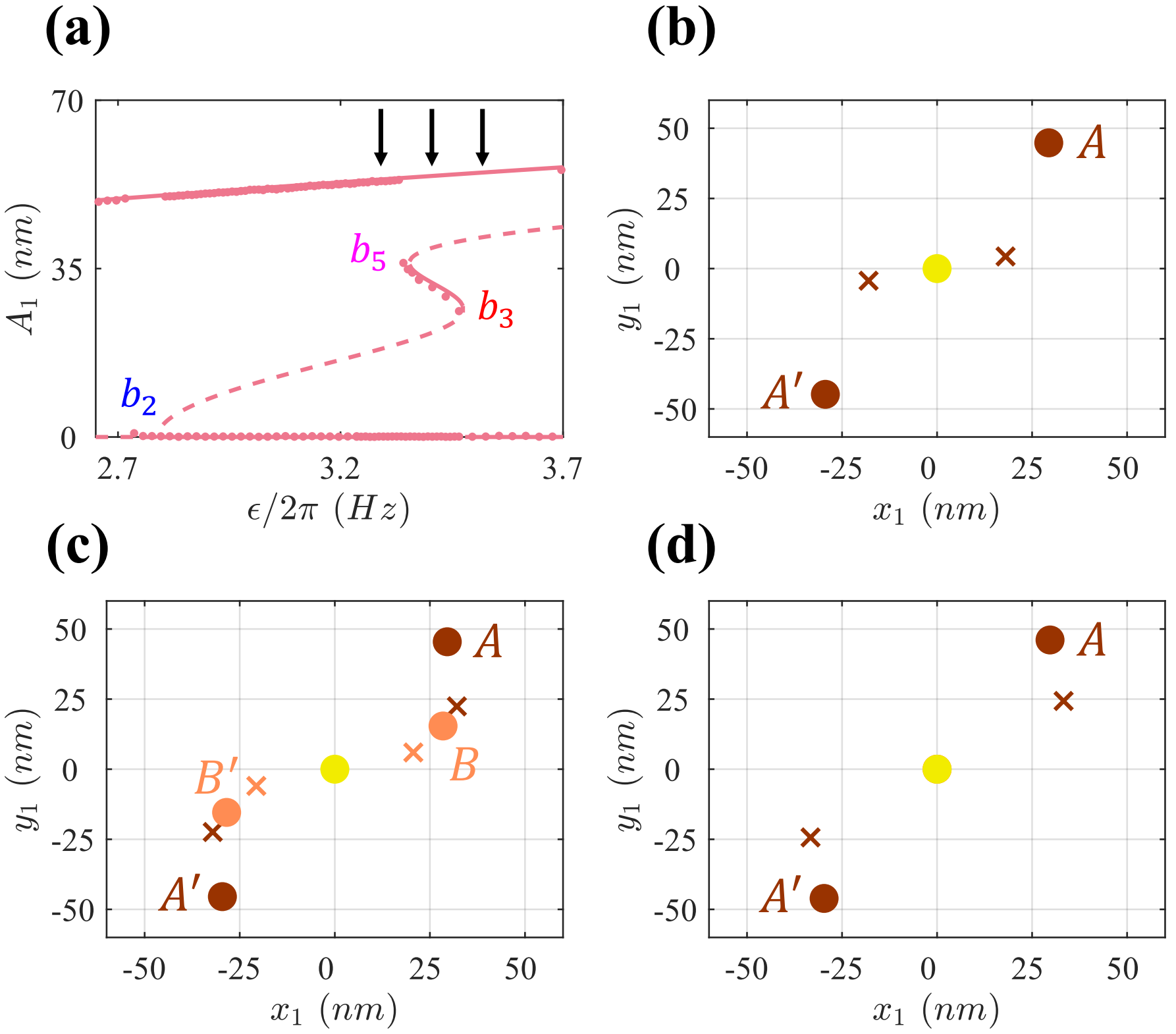

**Supplementary Figure 3.** Emergence of second pair of period-two states on the phase plane of the quadratures of period-two vibrations. $A$ and $A'$, $B$ and $B'$ (full circles) indicate stable period-two states. Crosses indicate unstable period-two states. These states correspond to stationary states in the rotating frame. (a) Frequency dependence of the vibration amplitude in the light blue regions of figure 4(b) of the main text. Dots show the measurements, solid lines show the theory. (b) Stable and unstable stationary states for the value of $\epsilon$ indicated by the left arrow in (a), $\epsilon < \epsilon_{b_5}$. The center corresponds to the stable zero-amplitude state. There is one pair of period-two states in this region. (c) For $\epsilon$ between its values at $b_5$ and $b_3$, the central arrow in panel (a), the oscillator has two pairs of stable period-two states. (d) For the value of $\epsilon$ indicated by the right arrow in panel (a), $\epsilon > \epsilon_{b_3}$, there is one pair of stable period-two states.